\documentclass[aps,prl,twocolumn,groupedaddress]{revtex4}

\usepackage{amsmath,amssymb}	
\usepackage{graphicx}

\begin{document}

\title{Radial Spreading of Drift Wave-Zonal Flow Turbulence via Soliton Formation}

\author{Zehua Guo}
\author{Liu Chen}
\altaffiliation{Institute for Fusion Theory and Simulation, ZheJiang University, China}
\affiliation{Department of Physics and Astronomy,  University of California, Irvine, CA 92697, U.S.A.}
\author{Fulvio Zonca}
\affiliation{Associazione EURATOM-ENEA sulla Fusione, C.P. 65--00044  Frascati, Italy}

\date{\today}

\begin{abstract}
The self-consistent spatiotemporal evolution of drift wave (DW) radial envelope and zonal flow (ZF) amplitude is investigated in a slab model~\cite{guzdar01}. Stationary solution of the coupled partial differential equations in a simple limit yields formation of DW-ZF soliton structures, which propagate at group velocity depending on the envelope peak amplitude. Additional interesting physics, {\it e.g.} generation, destruction, collision and reflection of solitons, as well as turbulence bursting can also be observed due to effects of linear growth/damping, dissipation, equilibrium nonuniformities and soliton dynamics. The propagation of soliton causes significant radial spreading of DW turbulence and therefore can affect transport scaling with system size by broadening of the turbulent region. Correspondence of the present analysis with the description of DW-ZF interactions in toroidal geometry~\cite{chen04, zonca04} is also elucidated.
\end{abstract}


\maketitle

Explaining the size scaling of confinement properties in magnetized plasmas is one of the crucial and challenging problems of fusion energy research. It has been pointed out that turbulence spreading is responsible for local turbulence intensity dependence on global equilibrium properties~\cite{lin02m}, {\it i.e.} the system size, and, thus, for the size scaling of turbulent transport coefficients. Therefore, the nonlocal character of turbulent intensity plays a crucial role in the breakdown of gyro-Bohm scaling of turbulent transport and transition to Bohm scaling, as observed in several numerical simulations~\cite{chen04,lin02,lin04}.

The radial propagation of drift-wave (DW) turbulence in tokamak plasmas was first investigated by Garbet~\cite{garbet94}, in the absence of zonal flow (ZF). Turbulence spreading was investigated also in Refs.~\cite{diamond95a,diamond95b}. Later on, using a single model equation for the local turbulence intensity, Hahm {\it et. al.}~\cite{hahm04} considered the ``minimal problem" for turbulence spreading, which is about spatiotemporal diffusive propagation of a patch of turbulence as a fluctuation front from an unstable to a stable or a weaker drive region. A mean field theory of turbulent transport has been developed and extensively studied. By performing a Fokker-Planck analysis on the evolution of turbulence energy density, or applying quasilinear theory to the wave kinetic equation, one can derive a simple equation for the mean turbulence energy density. This approach leads to a reaction diffusion equation similar to the well-known Fisher equation~\cite{fisher37,kolmogorov37}. G\"urcan {\it et al.}~\cite{gurcan05} obtained an exact solution for this model, which describes a ballistic front propagation with speed given by the geometric mean of diffusion coefficient and linear growth. In this work it was pointed out that ballistic spreading is possible even without toroidal coupling effects. A more systematic approach~\cite{gurcan07} was proposed to explain turbulence spreading in terms of nonlinear mode couplings using a two field Hasegawa-Wakatani model (kinetic and internal energy) recovering the previous one-field model~\cite{gurcan05} in the proper limit, where the fluxes due to nonlinear interaction are written in the Fick's law form. Analyses of turbulence spreading based on solutions of a bi-variate Burgers equation~\cite{burgers48} for the evolution of the DW plasmon density  were reported in Ref.~\cite{kim03}. Garbet {\it et al.}~\cite{garbet07} also developed a two-field critical gradient model that couples a heat equation to an evolution equation for the turbulence intensity. It is shown that this model exhibits the dual character of turbulent dynamics, diffusive or ballistic, depending on parameters such as the heat flux and the wave number.

In spite of great efforts, the fundamental dynamics of turbulence spreading is still not well understood. Although turbulence is truly a microscopic phenomenon, spreading or propagation of turbulence is usually related to mesoscale dynamics, {\it e.g.} intermittency, formation of avalanches, transport barriers and other coherent structures, which cannot be described by linear excitation and nonlinear wave-wave couplings via triad interaction processes only. ZFs are frequently assumed to be less or not important at all in the spreading process~\cite{hahm04,gurcan05,gurcan07}, based on the argument that large scale radially extended eddies are most effective at spreading turbulence, while ZFs inhibit spreading by destroying these structures~\cite{gurcan06,gurcan06b}. However, slower DW turbulence spreading, observed in global gyrokinetic simulations when ZFs are included, has been attributed to the suppression of DW intensity by the ZFs~\cite{wang06,wang07} and not to their dynamic role.

In the present work, we study the nonlinear DW-ZF interplay in a simple slab geometry in order to elucidate the underlying physics mechanisms responsible for turbulence spreading. A general two-field DW-ZF model is derived for the spatiotemporal evolution of the DW radial envelope and ZF amplitude, which reduces to previous descriptions~\cite{chen04,zonca04,chen00} when ZF induced  modulations on a given DW pump are considered with its sidebands (4-wave). Since the total energy cascades into shorter radial wavelengths via the coherent nonlinear DW-ZF modulation interaction, the local DW envelope nonlinearly steepens and the DW linear dispersion becomes stronger. Time scales for nonlinear interaction and linear dispersion eventually become comparable, showing analogies to the Langmuir soliton problem. Coherent structures are, thus, expected to form, such as DW-ZF solitons which will propagate radially. Turbulence spreading may then occur via DW-ZF soliton propagation with $x\sim t$,  which is  faster than any diffusive process. 
\begin{figure}[t]
\includegraphics{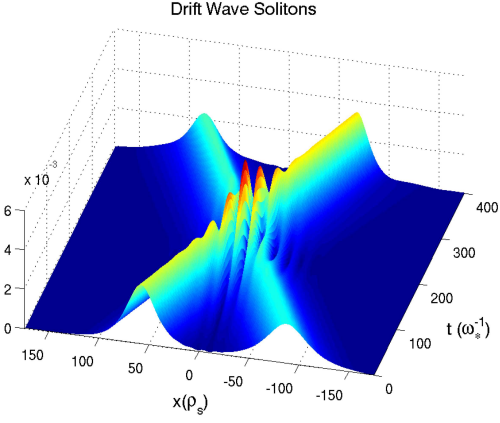}%
\caption{\label{2soliton}Collision between two different DW-ZF solitons, where $A_{d1} = 0.002$, $k_x\rho_s = 0.3$ and $A_{d2}=0.0003$, $k_x\rho_s = 0.3$.}
\end{figure}

The coherent 4-wave DW-ZF modulational interaction model~\cite{chen00} has been applied to study turbulence spreading in toroidal plasmas, demonstrating that coupled nonlinear evolution equations for DW radial envelope and ZF structures can generally be derived from first principles~\cite{chen04, zonca04}. In this work, the same approach is applied in a simplified slab geometry~\cite{guzdar01}, where $x, y, z$ corresponds to toroidal coordinates $r, \theta, \zeta$, respectively, and the radial wave number $k_x$ is equivalent to $k_r \equiv n \theta_k dq/dr$~\cite{chen04,zonca04,chen00} of the DW radial envelope. This simplified approach also helps elucidating ``the subtle differences between the slab and toroidal geometries''~\cite{guzdar01} and yields nonlinear equations that are derived from first principles as those of~\cite{chen04,zonca04}. This important point allows us to support the generality of our results reported hereafter, which do not rely on any ad-hoc model assumption for the description of nonliner DW-ZF interplay.

We start from the slab analysis of the electrostatic DW-ZF interaction model proposed in~\cite{guzdar01}. Similar to the Hasegawa-Mima's model, using two-fluid description and quasi-neutrality condition, one can straightforwardly derive the DW evolution equation in the form~\cite{guzdar01}:
\begin{eqnarray}\label{eq:dw}
\lefteqn{(1-\rho_s^2 \nabla^2)\partial_t \phi_d - (c_s^2/\Omega_i) \nabla \phi_d \times {\hat{z}}\cdot \nabla \ln n_0 }  \nonumber \\ & & - (c_s^2/\Omega_i) \nabla \phi_z \times {\hat{z}}\cdot \nabla \phi_d + (c_s^2 \rho_s^2/\Omega_i) \nabla \cdot \left[ \nabla\phi_z \times {\hat{z}}\right. \nonumber  \\ & &  \left. \cdot \nabla \nabla\phi_d  + \nabla \phi_d \times {\hat{z}}\cdot \nabla \nabla\phi_z\right] = 0\,;
\end{eqnarray}
where $c_s = \sqrt{T_e / m_i}$, $\rho_s = c_s / \Omega_i$ is the ion Larmor radius at the sound speed and the scalar potential is normalized to $T_e/e$; the ZF potential $\phi_z = \langle\phi\rangle$, where $\langle\cdots\rangle$ represents flux surface averaging ($(y, z)$ plane). The last two terms on the right-hand side correspond to higher order Reynolds stress corrections due to nonlinear polarization drift, ${\rm O}(k^4\rho_s^4)$, which can be ignored when $\vert k\rho_s\vert \ll 1$. Meanwhile, ZF has $k_\theta=k_\parallel=0$; thus, electrons do not behave adiabatically in the ZF potential. We can describe ZFs by the condition of no net radial flux:
\begin{equation}\label{eq:zf}
\partial_t \nabla^2 \phi_z - (c_s^2/\Omega_i) \langle\nabla \cdot [ \nabla \phi_d \times {\hat{z}}\cdot \nabla \nabla \phi_d]\rangle = 0\,.
\end{equation}
As in Refs.~\cite{chen04, guzdar01, zonca04}, we consider a coherent drift wave with single toroidal number $n$, or constant $k_y$ in slab geometry. Thus, the 2-field coupled set of DW-ZF evolution equations are readily cast in the form
\begin{align}
(1 + k_y^2 - \partial_x^2 ) \partial_t \phi_d + i \omega_\ast (x) \phi_d = - i C \phi_d \partial_x \phi_z & \label{eq:model_1}\,,\\
\partial_t\phi_z = i C \langle\phi_d \partial_x \phi_d^\ast  - {\it c.c.}\rangle &\label{eq:model_2}\,;
\end{align}
where  $\omega_\ast = - k_y (c_s^2/\Omega_i) d \ln n/dx$ is the diamagnetic drift frequency, $C = k_y \Omega_i/\omega_\ast$ is a constant, while space and time have been normalized to $\rho_s$ and $\omega_\ast^{-1}$ respectively. Note the structural similarity of Eqs.\eqref{eq:model_1}-\eqref{eq:model_2} with Eqs.~(4) of~\cite{chen04} in toroidal geometry. Numerical simulations of the above coupled system, given $\phi_d$ as a Gaussian function of $x$ at $t=0$, show that DW-ZF can form solitary structures, which coherently propagate with given group velocity (Fig.1). These coherent structures are envelope solitons with wavelength of the carrier wave comparable to the envelope width, suggesting that turbulence spreading can be caused by soliton formation due to balance between DW dispersion and trapping by nonlinearly generated ZFs.

For the sake of simplicity, we initially ignore linear growth/damping and dissipation of both DW and ZF. For now, we also take $\omega_\ast$ constant. The  $\omega_\ast(x)$ profile introduces extra effects of finite system size, which will be discussed elsewhere. Furthermore, we assume a coherent DW form $\phi_d(x,t) = A_d u_d(x,t) {\rm exp}(i k_x x - i \omega t)$, in which $A_d$ is the maximum perturbation amplitude, usually  $\approx 10^{-4}-10^{-2}$, the normalized envelope function $u_d(x,t)$ is chosen to be real and long-scale $\vert\partial_x^2{u_d}\vert \ll k_y^2 \vert u_d\vert$, the phase $\varphi = k_x x - \omega t$ describes fast oscillations in time but not necessarily in space, $k_x$ is the radial wave number and $\omega$ is the DW frequency. Once the given DW form is substituted into Eqs.\eqref{eq:model_1}-\eqref{eq:model_2}, the coupled PDEs can be rewritten in the form of a nonlinear Schr\"odinger equation
\begin{eqnarray}
& &(1+k_\perp^2)(\partial_t + v_g\partial_x) u_d + (i\omega - \partial_t)\partial_x^2 u_d \nonumber \\ & & \hspace*{2em} - i \omega^2\lambda u_d = -i C\partial_x\phi_z u_d \label{eq:schrod_1}\,,\\
& &\partial_t \phi_z = 2Ck_xA_d^2u_d^2\label{eq:schrod_2}\,;
\end{eqnarray}
where $v_g = - 2k_x\omega/(1+k_\perp^2)$, $\lambda=(1+k_\perp^2)-1/\omega$, and $k_\perp^2 = k_x^2+k_y^2$. For constructing a stationary solution, we introduce $\xi = (x-v_gt)\delta$ and  then assume $u_d(x,t) = u_d(\xi)$ and $\phi_z(x,t) = \phi_z(\xi)$, such that $\partial_t = - v_g\delta\partial_\xi$ and $\partial_x = \delta\partial_\xi$. Here, the small parameter $\delta$ corresponds to the slowly varying envelope scale. Finally, substituting $\phi_z$ from Eq.\eqref{eq:schrod_2} into Eq.\eqref{eq:schrod_1}, we derive one single ordinary differential equation (ODE) for the DW perturbation: 
\begin{equation}\label{eq:ode_dim}
\delta^2u_d'' - \lambda u_d + (C^2/\omega^2)(1+k_\perp^2)A_d^2u_d^3 = 0\,;
\end{equation}
where terms $\propto \partial_t + v_g\partial_x$ cancel by construction and $\partial_x^2\partial_t u_d$ is ignored assuming that the envelope transient time is much longer than the DW oscillation period, {\it e.g.} $v_g\delta\ll\omega$, which can be justified {\it a posteriori}. The above second order ODE clearly indicates the competition between linear dispersion and nonlinear self-trapping process. When the DW amplitude $A_d$ increases, its envelope becomes nonlinearly steeper, {\it i.e.} $\delta$ increases; meanwhile, the DW dispersion also becomes stronger and tends to inhibit the focusing process. Formally, this corresponds to equating the three coefficients of $u_d''$, $u_d$ and $u_d^3$, {\it i.e.}
\begin{equation}\label{eq:3coef}
\delta^2 = 1+k_\perp^2- 1/\omega = C^2/(2\omega^2)(1+k_\perp^2)A_d^2
\end{equation}
The DW wave-packet frequency $\omega$ is then readily obtained from the above quadratic equation, {\it i.e.} $\omega = \big[ 1 + \sqrt{1+2(1+k_\perp^2)^2 C^2 A_d^2}\big]/\big[2(1+k_\perp^2)\big]$. Note that the right-hand side contains the nonlinear frequency shift due to finite DW turbulence amplitude. Similarly, the parameter $\delta$ can also be determined as 
\begin{equation}\label{eq:delta}
\delta = \pm\sqrt{(1+k_\perp^2)/2}C(A_d/\omega)
\end{equation}
\begin{figure}[t]
\includegraphics{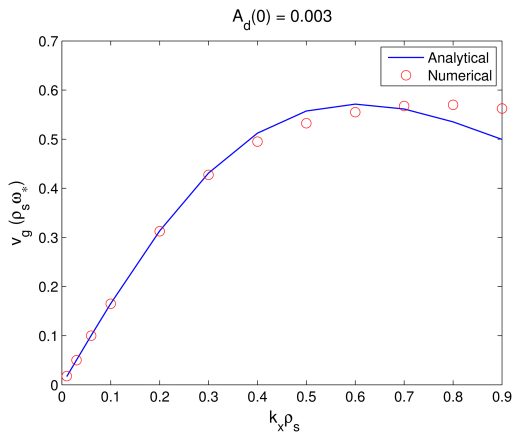}%
\caption{\label{vgkx}The relation between $v_g$ and $k_x$ for $A_d = 0.003$.}
\end{figure}
This derivation is subject to our  {\it a priori} assumption $\delta\ll\omega v_g^{-1}$, which guarantees that the DW oscillation $\omega$ occurs on the fastest timescale. Substituting $\delta, \omega, v_g$ as functions of $k_x$ and $A_d$,  this assumption is equivalent to $k_xk_y\Omega_i/\omega_\ast A_d \ll \sqrt{(1+k_\perp^2)/2}\,\omega(k_x, A_d)$. Finally, given Eq.\eqref{eq:3coef}, Eq.\eqref{eq:ode_dim} becomes a dimensionless ODE governing the stationary envelope function $u_d$,
\begin{equation}\label{eq:ode_ndim}
u_d'' - u_d + 2 u_d^3 = 0 .
\end{equation}
This ODE is analogous to that of  an oscillator in the so called ``Sagdeev potential" $\Phi(u_d) = (-u_d^2 + u_d^4)/2$, whose solution can be written as hyperbolic secant function $u_d(\xi) = {\rm Sech} (\xi)$, when appropriate boundary conditions are imposed, {\it viz}. $u_d\rightarrow 0$ at $\vert\xi\vert\rightarrow\infty$. Meanwhile, the ZF solution is obtained straightforwardly by integrating Eq.\eqref{eq:schrod_2} once, such that $\phi_z(\xi) = \int 2k_x C/(v_g\delta)A_d^2{\rm Sech}^2(\xi)d\xi = \mp\sqrt{2(1+k_\perp^2)}\,A_d{\rm Tanh}(\xi)$ which statisfies the causality constraint, {\it i.e.} $\partial_\xi\phi_z\rightarrow 0$ when $\vert\xi\vert\rightarrow\infty$ for any initially localized DW turbulence. The expressions for DW and ZF in the laboratory frame are
\begin{align}
&\phi_d(x,t) = A_d {\rm Sech}\big[\delta(x+\dfrac{2k_x\omega}{1+k_\perp^2}t)\big]  e^{ik_xx-i\omega t} \label{eq:sol_dw}\,,\\
&\phi_z(x,t) =\sqrt{2(1+k_\perp^2)}A_d{\rm Tanh}\big[\delta(x+\dfrac{2k_x\omega}{1+k_\perp^2}t)\big]\label{eq:sol_zf}\,.
\end{align}
%
From Eqs.\eqref{eq:sol_dw} and \eqref{eq:sol_zf}, it generally follows that ZF potentials have radially moving structures of hyperbolic tangent shape; meanwhile, $E_z = - \partial\phi_z/\partial_x$ manifests itself as scalar-potential wells in the background plasma and trap the corresponding DW packets. Figure~\ref{2soliton} shows the spatiotemporal evolution of two counter-propagating DW-ZF solitons, which are solutions of the original coupled PDEs, given $k_y=0.3$, $\Omega_i/\omega = 100$. For consistency with our analytic approach, we have chosen initial $k_x$ and $A_d$ to satisfy the a priori assumption $\delta\ll\omega v_g^{-1}$. Furthermore, we assumed no $\omega_\ast$ equilibrium variation and no growth/damping and dissipation. Note that the two envelope solitons remain unchanged in both real and $k$ space after the collision, although the dynamics during the collision can be quite complicated. This is one of the soliton essential features.
\begin{figure}[t]
\includegraphics{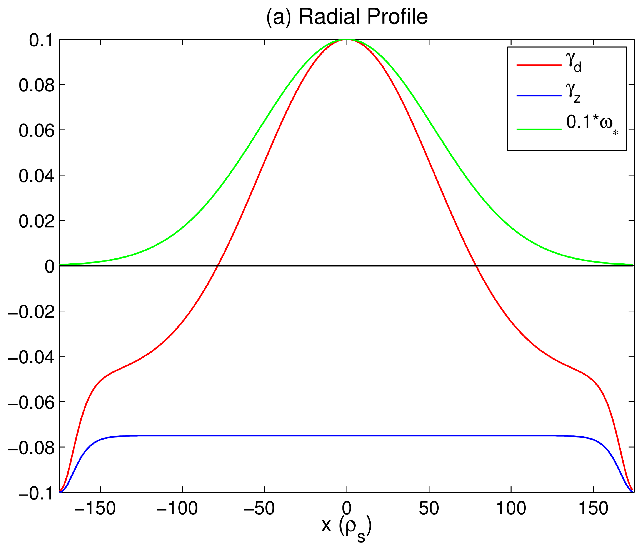}\\%
\includegraphics{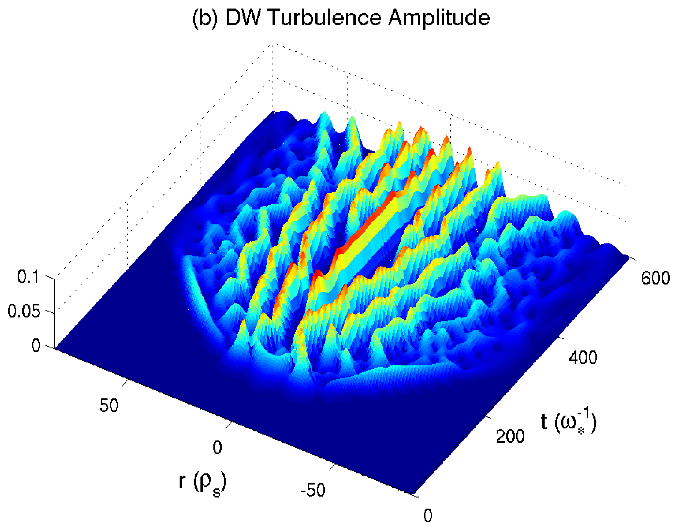}%
\caption{\label{turb}DW turbulence spreading in the presence of growth/damping, dissipation and finite system size effects.}
\end{figure}

The radial propagation velocity of DW-ZF solitary structures, $v_g$, depends on both the radial wave number and the DW amplitude. It is different from the linear group velocity, $v'_g=\partial\omega_l/\partial k_x$, which is determined by $k_x$ through the linear dispersion relation only. Therefore the solution of Eqs.\eqref{eq:sol_dw}-\eqref{eq:sol_zf} gives a two-parameter, $k_x$ and $A_d$, family of solitons. Figure~\ref{vgkx} shows the relation between $v_g$ and $k_x$ for small initial amplitude $A_d = 0.003$. Numerical and analytical results agree well when $k_x \lesssim 0.4$. The discrepancy  when $k_x > 0.4$ originates from the breaking of the a priori assumption that $\partial_x^2\partial_t\phi_d$ can be ignored in Eq.\eqref{eq:schrod_1}. Moreover, when $A_d$ increases to about $10^{-2}$, the analytical result is no longer valid either, since the ignored term ${\rm O}(v_g\delta^3)$ modifies the solution at larger $k_x$ or $A_d$, according to Eq.\eqref{eq:delta}. 

Our numerical simulation results for $A_d\gtrsim 0.01$ show that the dominant asymptotic ($t\rightarrow\infty)$  DW turbulence behavior is still of soliton type and the propagation velocity $v_g$ increases with the DW amplitude $A_d$. We observe that the DW radial wave number no longer corresponds to its initial value but rather to $\delta$, which is mainly determined by the amplitude $A_d$ alone. There seems to be a transition from a 2-parameter to a 1-parameter family of soliton solutions of the coupled system. If the amplitude becomes even larger, {\it e.g.} $A_d\geqslant 0.02$, the initially localized DW-ZF soliton breaks into many pieces in the form of solitons and wave-trains; similar to Gardner's work~\cite{gardner67} on Korteweg-deVries equation, in which it is shown that a localized but otherwise arbitrary initial perturbation will generate a conventional wave-train, quickly destroyed by dispersion, and a finite number of solitons, which characterize the asymptotic solution. 
 
We studied the DW-ZF initial value problem in more general cases as well, {\it i.e.} in the presence of linear growth/damping, dissipations, and variation of equilibrium profiles. Figure~\ref{turb} shows the evolution of DW turbulence out of initial random noise, with strong DW growth rate $\gamma_d(0) = 0.1$, uniform ZF damping rate $\gamma_z = 0.075$ and $L_p = 150$, which represents the system size. Dissipations are also included. The drift frequency $\omega_\ast(x)$ has Gaussian shape centered at $x=0$ (Fig.~\ref{turb}(a)); DW turbulence is linearly unstable in the central region ($\vert x\vert<80\rho_s$) but is damped in the outer region ($\vert x\vert>80\rho_s$), while the ZF is uniformly damped. Figure~\ref{turb}(b) clearly shows formation and propagation of solitons, which however exhibit more complicate dynamic behaviors; for instance, growing amplitudes, slowing down of propagation speed, soliton breaking, turbulence bursting and more. Since coupled PDEs generally describe infinite-dimensional dynamical systems, DW turbulence dynamics appears mostly chaotic in the corresponding parameter space, $(\gamma_d, \gamma_z)$. Solitons may bounce back at their turning points, possibly enhancing nonlinear interactions inside the turbulent region and impacting the size scaling of turbulence transport. Figure~\ref{turb}(b) also demonstrates that the nonlinearly saturated turbulence has spread into a much broader region than that of its linear mode structure. 

In summary, we have demonstrated the novel result that coherent structures such as solitons can be constructed self-consistently in a two-field DW-ZF model and cause significant radial turbulence spreading in a slab plasma. We have also shown the structural analogy of the underlying coupled PDEs for the nonlinear evolution of DW radial envelope and ZF amplitude with the corresponding equations derived in toroidal geometry~\cite{chen04,zonca04}, demonstrating the generality of the present results and the possibility of readily extending them in future works. The size scaling of DW turbulence will also be discussed in detail in a separate work.

This work was supported by U.S. DOE and NSF grants. The authors acknowledge useful discussions with Z. Lin.


\begin{thebibliography}{23}
\expandafter\ifx\csname natexlab\endcsname\relax\def\natexlab#1{#1}\fi
\expandafter\ifx\csname bibnamefont\endcsname\relax
  \def\bibnamefont#1{#1}\fi
\expandafter\ifx\csname bibfnamefont\endcsname\relax
  \def\bibfnamefont#1{#1}\fi
\expandafter\ifx\csname citenamefont\endcsname\relax
  \def\citenamefont#1{#1}\fi
\expandafter\ifx\csname url\endcsname\relax
  \def\url#1{\texttt{#1}}\fi
\expandafter\ifx\csname urlprefix\endcsname\relax\def\urlprefix{URL }\fi
\providecommand{\bibinfo}[2]{#2}
\providecommand{\eprint}[2][]{\url{#2}}

\bibitem[{\citenamefont{{Guzdar {\sl et al.}}}(2001)}]{guzdar01}
\bibinfo{author}{\bibfnamefont{P.~N.} \bibnamefont{{Guzdar {\sl et al.}}}},
  \bibinfo{journal}{Phys. Plasmas} \textbf{\bibinfo{volume}{8}},
  \bibinfo{pages}{459} (\bibinfo{year}{2001}).

\bibitem[{\citenamefont{{Chen {\sl et al.}}}(2004)}]{chen04}
\bibinfo{author}{\bibfnamefont{L.}~\bibnamefont{{Chen {\sl et al.}}}},
  \bibinfo{journal}{Phys. Rev. Lett.} \textbf{\bibinfo{volume}{92}},
  \bibinfo{pages}{075004} (\bibinfo{year}{2004}).

\bibitem[{\citenamefont{{Zonca {\sl et al.}}}(2004)}]{zonca04}
\bibinfo{author}{\bibfnamefont{F.}~\bibnamefont{{Zonca {\sl et al.}}}},
  \bibinfo{journal}{Phys. Plasmas} \textbf{\bibinfo{volume}{11}},
  \bibinfo{pages}{2488} (\bibinfo{year}{2004}).

\bibitem[{\citenamefont{{Lin {\sl et al.}}}(2002{\natexlab{a}})}]{lin02m}
\bibinfo{author}{\bibfnamefont{Z.}~\bibnamefont{{Lin {\sl et al.}}}},
  \bibinfo{journal}{Proc. 19th Int. Conf. on Plasma Phys. and Control. Nuc.
  Fusion Res., Lyon, France, 2002 (IAEA, Vienna, 2002).} pp.
  \bibinfo{pages}{IAEA--CN--94/TH1/1} (\bibinfo{year}{2002}{\natexlab{a}}).

\bibitem[{\citenamefont{{Lin {\sl et al.}}}(2002{\natexlab{b}})}]{lin02}
\bibinfo{author}{\bibfnamefont{Z.}~\bibnamefont{{Lin {\sl et al.}}}},
  \bibinfo{journal}{Phys. Rev. Lett.} \textbf{\bibinfo{volume}{88}},
  \bibinfo{pages}{195004} (\bibinfo{year}{2002}{\natexlab{b}}).

\bibitem[{\citenamefont{Lin and Hahm}(2004)}]{lin04}
\bibinfo{author}{\bibfnamefont{Z.}~\bibnamefont{Lin}} \bibnamefont{and}
  \bibinfo{author}{\bibfnamefont{T.~S.} \bibnamefont{Hahm}},
  \bibinfo{journal}{Phys. Plasmas} \textbf{\bibinfo{volume}{11}},
  \bibinfo{pages}{1099} (\bibinfo{year}{2004}).

\bibitem[{\citenamefont{{Garbet {\sl et al.}}}(1994)}]{garbet94}
\bibinfo{author}{\bibfnamefont{X.}~\bibnamefont{{Garbet {\sl et al.}}}},
  \bibinfo{journal}{Nucl. Fusion} \textbf{\bibinfo{volume}{34}},
  \bibinfo{pages}{963} (\bibinfo{year}{1994}).

\bibitem[{\citenamefont{{Diamond {\sl et al.}}}(1995)}]{diamond95a}
\bibinfo{author}{\bibfnamefont{P.~H.} \bibnamefont{{Diamond {\sl et al.}}}},
  \bibinfo{journal}{Phys. Plasmas} \textbf{\bibinfo{volume}{2}},
  \bibinfo{pages}{3685} (\bibinfo{year}{1995}).

\bibitem[{\citenamefont{Diamond and Hahm}(1995)}]{diamond95b}
\bibinfo{author}{\bibfnamefont{P.~H.} \bibnamefont{Diamond}} \bibnamefont{and}
  \bibinfo{author}{\bibfnamefont{T.~S.} \bibnamefont{Hahm}},
  \bibinfo{journal}{Phys. Plasmas} \textbf{\bibinfo{volume}{2}},
  \bibinfo{pages}{3640} (\bibinfo{year}{1995}).

\bibitem[{\citenamefont{{Hahm, P. H. Diamond, Z. Lin {\sl et
  al.}}}(2004)}]{hahm04}
\bibinfo{author}{\bibfnamefont{T.~S.} \bibnamefont{{Hahm, P. H. Diamond, Z. Lin
  {\sl et al.}}}}, \bibinfo{journal}{Plasma Phys. Control. Fusion}
  \textbf{\bibinfo{volume}{46}}, \bibinfo{pages}{A323} (\bibinfo{year}{2004}).

\bibitem[{\citenamefont{Fisher}(1937)}]{fisher37}
\bibinfo{author}{\bibfnamefont{R.~A.} \bibnamefont{Fisher}},
  \bibinfo{journal}{Ann. Eugenics} \textbf{\bibinfo{volume}{7}},
  \bibinfo{pages}{353} (\bibinfo{year}{1937}).

\bibitem[{\citenamefont{Kolmogorov et~al.}(1937)\citenamefont{Kolmogorov,
  Petrovskii, and Piscounov}}]{kolmogorov37}
\bibinfo{author}{\bibfnamefont{A.}~\bibnamefont{Kolmogorov}},
  \bibinfo{author}{\bibfnamefont{I.}~\bibnamefont{Petrovskii}},
  \bibnamefont{and}
  \bibinfo{author}{\bibfnamefont{N.}~\bibnamefont{Piscounov}},
  \bibinfo{journal}{Bull. Moscow Univ., Math. Mech.}
  \textbf{\bibinfo{volume}{1}}, \bibinfo{pages}{1} (\bibinfo{year}{1937}).

\bibitem[{\citenamefont{{G\"urcan {\sl et al.}}}(2005)}]{gurcan05}
\bibinfo{author}{\bibfnamefont{O.~D.} \bibnamefont{{G\"urcan {\sl et al.}}}},
  \bibinfo{journal}{Phys. Plasmas} \textbf{\bibinfo{volume}{12}},
  \bibinfo{eid}{032303} (\bibinfo{year}{2005}).

\bibitem[{\citenamefont{{G\"urcan {\sl et al.}}}(2007)}]{gurcan07}
\bibinfo{author}{\bibfnamefont{O.~D.} \bibnamefont{{G\"urcan {\sl et al.}}}},
  \bibinfo{journal}{Phys. Plasmas} \textbf{\bibinfo{volume}{14}},
  \bibinfo{eid}{055902} (\bibinfo{year}{2007}).

\bibitem[{\citenamefont{Burgers}(1948)}]{burgers48}
\bibinfo{author}{\bibfnamefont{J.~M.} \bibnamefont{Burgers}},
  \bibinfo{journal}{Adv. Appl. Mech} \textbf{\bibinfo{volume}{1}},
  \bibinfo{pages}{171} (\bibinfo{year}{1948}).

\bibitem[{\citenamefont{{Kim {\sl et al.}}}(2003)}]{kim03}
\bibinfo{author}{\bibfnamefont{E.~J.} \bibnamefont{{Kim {\sl et al.}}}},
  \bibinfo{journal}{Nucl. Fusion} \textbf{\bibinfo{volume}{43}},
  \bibinfo{pages}{961} (\bibinfo{year}{2003}).

\bibitem[{\citenamefont{{Garbet {\sl et al.}}}(2007)}]{garbet07}
\bibinfo{author}{\bibfnamefont{X.}~\bibnamefont{{Garbet {\sl et al.}}}},
  \bibinfo{journal}{Phys. Plasmas} \textbf{\bibinfo{volume}{14}},
  \bibinfo{eid}{122305} (\bibinfo{year}{2007}).

\bibitem[{\citenamefont{{G\"urcan {\sl et
  al.}}}(2006{\natexlab{a}})}]{gurcan06}
\bibinfo{author}{\bibfnamefont{O.~D.} \bibnamefont{{G\"urcan {\sl et al.}}}},
  \bibinfo{journal}{Phys. Rev. Lett.} \textbf{\bibinfo{volume}{97}},
  \bibinfo{eid}{024502} (\bibinfo{year}{2006}{\natexlab{a}}).

\bibitem[{\citenamefont{{G\"urcan {\sl et
  al.}}}(2006{\natexlab{b}})}]{gurcan06b}
\bibinfo{author}{\bibfnamefont{O.~D.} \bibnamefont{{G\"urcan {\sl et al.}}}},
  \bibinfo{journal}{Phys. Plasmas} \textbf{\bibinfo{volume}{13}},
  \bibinfo{eid}{052306} (\bibinfo{year}{2006}{\natexlab{b}}).

\bibitem[{\citenamefont{{Wang {\sl et al.}}}(2006)}]{wang06}
\bibinfo{author}{\bibfnamefont{W.~X.} \bibnamefont{{Wang {\sl et al.}}}},
  \bibinfo{journal}{Phys. Plasmas} \textbf{\bibinfo{volume}{13}},
  \bibinfo{pages}{092505} (\bibinfo{year}{2006}).

\bibitem[{\citenamefont{{Wang {\sl et al.}}}(2007)}]{wang07}
\bibinfo{author}{\bibfnamefont{W.~X.} \bibnamefont{{Wang {\sl et al.}}}},
  \bibinfo{journal}{Phys. Plasmas} \textbf{\bibinfo{volume}{14}},
  \bibinfo{pages}{072306} (\bibinfo{year}{2007}).

\bibitem[{\citenamefont{{Chen {\sl et al.}}}(2000)}]{chen00}
\bibinfo{author}{\bibfnamefont{L.}~\bibnamefont{{Chen {\sl et al.}}}},
  \bibinfo{journal}{Phys. Plasmas} \textbf{\bibinfo{volume}{7}},
  \bibinfo{pages}{3129} (\bibinfo{year}{2000}).

\bibitem[{\citenamefont{{Gardner {\sl et al.}}}(1967)}]{gardner67}
\bibinfo{author}{\bibfnamefont{C.~S.} \bibnamefont{{Gardner {\sl et al.}}}},
  \bibinfo{journal}{Phys. Rev. Lett.} \textbf{\bibinfo{volume}{19}},
  \bibinfo{pages}{1095} (\bibinfo{year}{1967}).

\end{thebibliography}

\end{document}